\documentclass{aa}  

\usepackage{graphicx}
\usepackage{txfonts}
\usepackage[version=3]{mhchem}
\usepackage[dvipsnames]{xcolor}
\usepackage{lscape}
\usepackage{ulem}

\usepackage{booktabs}
\usepackage{hyperref}

\usepackage[figuresright]{rotating}
\usepackage{mathtools}
\usepackage{float}
\usepackage{csquotes}

\newcommand{\micron}{\textmu m\xspace}

\begin{document}

   \title{Chemistry and ro-vibrational excitation of HeH$^+$ \\ in the Planetary Nebula NGC 7027}

   \author{M. Sil\inst{\ref{inst1},\ref{inst2}} \and A. Faure\inst{\ref{inst1}} \and H. Wiesemeyer\inst{\ref{inst4}} \and P. Hily-Blant\inst{\ref{inst1}} \and J. Loreau\inst{\ref{inst3}} \and K.D. P\'{e}rez\inst{\ref{inst3}} \and {R. \v{C}ur\'{i}k} \inst{\ref{inst5}} \and F. Lique\inst{\ref{inst2}}}

   \institute{Univ. Grenoble Alpes, CNRS, IPAG, 38000 Grenoble, France\\
              \email{milansil93@gmail.com, alexandre.faure@univ-grenoble-alpes.fr}
              \label{inst1}
         \and
             Univ Rennes, CNRS, IPR (Institut de Physique de Rennes) - UMR 6251, F-35000 Rennes, France
             \label{inst2}
         \and
             Max-Planck-Institut für Radioastronomie, Auf dem Hügel 69, D-53121 Bonn, Germany
             \label{inst4}      
         \and
             KU Leuven, Department of Chemistry, B-3001 Leuven, Belgium
             \label{inst3}              
          \and
             J.~Heyrovsk\'{y} Institute of Physical Chemistry, Academy of Sciences, Dolej\v{s}kova~2155/3, 182~23 Prague~8, Czech Republic
             \label{inst5}}


 
  \abstract 
{The helium hydride cation (HeH$^+$) holds the distinction of being the first molecule to form in the metal-free Universe after the Big Bang. Following its first circumstellar detection via the pure rotational $J=1-0$ transition in the young and dense planetary nebula, NGC~7027, its presence was further confirmed by detecting the $\upsilon=1-0 \ P(1)$ and $P(2)$ rovibrational line emissions.}
{HeH$^+$ belongs to the class of \enquote{reactive} ions that can be destroyed so quickly that
chemical formation and destruction rates may compete with inelastic rates and should be considered when solving the statistical equilibrium equations. This so-called chemical \enquote{pumping} or \enquote{excitation} effect is investigated here for the first time in HeH$^+$.}
{The chemical evolution of HeH$^+$ in NGC~7027 is modeled with the \textsc{Cloudy} photoionization code using updated reaction rate coefficients. The electron temperature and atomic/molecular densities are modeled as a function of position in the nebula. The non-local thermodynamic equilibrium (NLTE) analysis of the three observed HeH$^+$ emission lines is then performed with the \textsc{Cloudy} and \texttt{RADEX} codes using an extensive set of spectroscopic and inelastic collisional data suitable for the specific high-temperature environment of NGC~7027. In a second approach, chemical formation and destruction rates of HeH$^+$ are implemented in \texttt{RADEX}. This code is combined with a Markov Chain Monte Carlo (MCMC) sampling (performed on the \texttt{RADEX}-parameters space) in order to extract the best-fit HeH$^+$ column density and physical conditions from the observed line fluxes.
}
{The \textsc{Cloudy} and \texttt{RADEX} NLTE results are found in good agreement, and they reproduce the observed HeH$^+$ line fluxes to within a factor of $2-5$, and the $\upsilon=1-0 \ P(2)/P(1)$ line ratio to better than 20\%. Agreement to better than a factor of 2.3 is obtained when including the reaction between He($2^3S$) and H as an additional source of HeH$^+$ in the chemical model.
The \texttt{RADEX/MCMC} model with chemical pumping is found to reproduce both the observed line fluxes and the line ratio to 20\%. {However, while the best-fit parameters agree rather well with the model predictions, the posterior distributions are poorly constrained,} suggesting that additional HeH$^+$ lines must be detected in NGC~7027 to better constrain the physical conditions via NLTE models. On the other hand, uncertainties in collisional (reactive and inelastic) data of HeH$^+$ have been largely reduced in this work.
We also show that the three observed lines are not sensitive to chemical pumping while excited \enquote{short-lived} levels are significantly overpopulated with respect to a NLTE model neglecting chemical excitation. The corresponding full-line spectrum predicted for NGC~7027 is provided.
}
   {}
   
\keywords{Astrochemistry -- Interstellar medium (ISM) -- ISM: molecules -- ISM: abundances -- Molecular processes -- Molecular data -- ISM: planetary nebulae: individual: NGC 7027}
   \maketitle
\section{Introduction \label{sec:intro}}

The formation of a planetary nebula begins when an intermediate-mass star sheds its dusty asymptotic giant branch (AGB) envelope at a high rate.
This stage is followed by the pre-planetary nebula phase, during which the central star, while still too cool to ionize its surroundings, drives fast energetic jets that shape the nebula.
Within a few hundred years, the fast-moving, mass-losing star sweeps the AGB material into a dense shell and ionizes it, completing the formation of the planetary nebula with a thick shell and a partly hollow interior.
NGC~7027 is a young, highly ionized, rapidly evolving, and extensively studied planetary nebula with a massive molecular envelope \citep{bubl23} within a distance $\sim1$~kpc of the Sun \citep[$D=980\pm100$~pc;][]{zijl08}.
The proper-motion analyses of the radio-frequency continuum emission from the inner elliptical shell indicate a kinematic age of $\sim600$~years \citep{mass89,zijl08}, whereas, based on multi-epoch HST imaging of the extended ring system, the nebular expansion analyses suggest a dynamical age between $1080-1620$~years \citep{scho18,guer20}.
However, both methods have inherent uncertainties, and the difference in ages derived from two different approaches suggests a multi-phase history for the nebula's formation and evolution.
The central star is exceptionally hot, with an effective temperature of about 198~kK and a significant bolometric luminosity of $\sim6.2\times10^3$~L$_{\sun}$ \citep{latt00,mora23}.
These parameters indicate that the central star originates from a progenitor of mass $\ge3$~M$_{\sun}$.
It has a present-day mass of $\sim0.6-0.7$~M$_{\sun}$ \citep{mora23}, and left the AGB phase within the last $\sim1000$~years \citep[from the stellar evolution models in][]{weid20}.
These extreme conditions lead to the formation of overlapping H and He Str\"{o}mgren spheres, which drive the production of the helium hydride cation (HeH$^+$, also known as hydro-helium cation) within a specific region where the He$^+$ Str\"{o}mgren sphere extends beyond the H$^+$ (H\,{\sc ii}) zone \citep{gust19,neuf20}.
{Although HeH$^+$ was initially identified in the laboratory a century ago \citep{hogn25}, the idea that it could exist in local astrophysical plasmas, particularly in planetary nebulae, was suggested more than four decades ago \citep{Dabrowski1977,blac78,Flower1979,robe82}.}

The near-infrared emissions from the $\upsilon = 1-0 \ P(1)$ and $P(2)$ ro-vibrational transitions of HeH$^+$ were recently detected by \cite{neuf20} in NGC~7027 planetary nebula using the iSHELL spectrograph on NASA’s Infrared Telescope Facility observations. This detection comes after the first-ever astrophysical identification of HeH$^+$ reported by \cite{gust19} using the GREAT instrument on the Stratospheric Observatory for Infrared Astronomy airborne observatory of its far-infrared $J = 1-0$ pure rotational transition toward the same target.
The confirmation of HeH$^+$ in nearby circumstellar space may provide important constraints on our understanding of the chemical network controlling its formation and destruction. The dominant formation channel of HeH$^+$ in NGC~7027 involves the radiative association of He$^+$ with neutral H \footnote{We note that in the early Universe, HeH$^+$ was formed by radiative association of neutral helium atoms with protons.} and the associative ionization of neutral H with He in the metastable level 2$^3$S state. The dominant destruction channels are the reaction of HeH$^+$ with H and electrons.

In agreement with earlier predictions, \cite{gust19} and \cite{neuf20} suggested that the prime region for the presence of HeH$^+$ (abundance peak) is near the Str\"{o}mgren radius of NGC~7027, where the physical conditions, such as electron density and temperature, abruptly change.
However, modeling the intensity of the observed HeH$^+$ lines has proven to be challenging.
A comprehensive nonequilibrium chemical and level population model, conducted by \cite{gust19}, predicted that the main-beam brightness temperature of the HeH$^+ \ J = 1-0$ line is approximately four times lower than the observed value. These authors suggested that this discrepancy could be resolved by adjusting the literature values of the rate coefficient for the radiative association process.
Later, \cite{neuf20} implemented several changes to the chemistry adopted by \cite{gust19} and, in particular, they found that the associative ionization between He($2^3S$) and H is another significant source of HeH$^+$. They predicted a line flux for the HeH$^+$ $\upsilon = 1-0 \ P(1)$ transition in very good agreement with the observed value. However, the measured fluxes of the $J = 1-0$ and $\upsilon = 1-0 \ P(2)$ lines were found to exceed the model predictions significantly by factors of 2.9 and 2.3, respectively. In addition, the observed $\upsilon=1-0 \ P(2)/P(1)$ line ratio ($1.34 \pm 0.24$) is significantly larger than the prediction (0.53).

In this work, we provide new rate coefficients for describing both the chemistry and excitation of HeH$^+$ in the specific high-temperature environment of NGC~7027. This new molecular data covers all 162 bound levels of HeH$^+$ and includes both state-to-state collisional inelastic (non-reactive) rate coefficients and state-resolved collisional reactive (formation and destruction) rate coefficients. The non-local thermodynamic equilibrium (NLTE) analysis of the observed HeH$^+$ rotational and ro-vibrational emissions is first performed with the \textsc{Cloudy} \citep{chat23} and \texttt{RADEX} \citep{vand07} codes. Both codes utilize our newly constructed HeH$^+$ collision data set for inelastic transitions.
In a second approach, chemical pumping of HeH$^+$ is implemented in \texttt{RADEX} and the observed line fluxes are fitted using Markov Chain Monte Carlo (MCMC) sampling \citep{Goodman2010} performed on the parameter space, namely the kinetic temperature, the atomic hydrogen and electron densities, and the HeH$^+$ column density.
In Sect.~\ref{sec:model}, we report the predicted results of our NLTE model and compare them with the observed HeH$^+$ line fluxes.
Finally, in Sect.~\ref{sec:discussion}, we discuss our obtained results and the concluding remarks are summarized.


\section{Modeling and results} \label{sec:model}

\subsection{\textsc{Cloudy} modeling and results}
\label{sec:cloudy}

The \textsc{Cloudy} photoionization code \citep[c23.01;][and references therein]{chat23} is used to model the HeH$^+$ abundance and interpret the observed line flux features in NGC~7027, similarly to \cite{gust19,neuf20}, by approximating an elongated nebula as spherically symmetric {and assuming dynamically young nebula in steady-state equilibrium}.

\subsubsection{Physical conditions} \label{sec:phy}

The physical model parameters are listed in \autoref{tab:phy}. Our standard model is computed under the assumption of constant pressure, as in \cite{neuf20}. The pressure is adjusted by fixing the initial total hydrogen nuclei density [$n({\rm H_{tot}})$] at $2.61\times10^4$~cm$^{-3}$ to obtain a Str\"{o}mgren sphere of angular radius at 4.6\arcsec ($\sim$ 0.02186~pc = $6.75\times10^{16}$~cm), considering a source distance of 980~pc.
{We note that the sound travel time ($\sim 1010$~years) in the constant pressure model is larger than the kinematic age \citep[$\sim 600$~years;][]{mass89,zijl08}, but comparable to the dynamic age \citep[$1080-1620$~years;][]{scho18,guer20} of the system.}
We assume the central star emits ionizing blackbody radiation with a temperature $1.98\times10^5$~K and a bolometric luminosity $6.2\times10^3$~L$_{\sun}$, irradiating a spherical geometry of gas \citep{latt00,mora23}.
The dimension of the gas is defined by an inner radius at 3.1\arcsec ($\sim$ 0.01473~pc = $4.55\times10^{16}$~cm).
The resulting geometry is a thick, spherical, and closed shell.
The ionization of atomic hydrogen by the Galactic cosmic-ray background is considered with a mean primary ionization rate of $\zeta_p = 2\times10^{-16}$~s$^{-1}$ \citep{indr07}. The ionization of H$_2$, with a mean secondary ionization rate of $\zeta_2 = 4.6\times10^{-16}$~s$^{-1}$ is estimated by the scaling $\zeta_2 = 2.3 \zeta_p$ \citep{glassgold1974}.
The background radiation field includes the cosmic microwave background (CMB) at $2.725$~K, the unextinguished local interstellar radiation field adopted from the continuum provided by \citet[see their Fig.~2]{blac87} and the Galactic background radiation field proposed by \cite{Draine1996} with an intensity of the continuum scaled up by a factor of $\sim 3\times10^4$ \citep{Hasegawa2000,Godard2013,Lau2016}.
The default elemental abundance (relative to total hydrogen nuclei) set for planetary nebulae in \textsc{Cloudy} modeling is primarily from \cite{alle83,khro89}.
Since NGC~7027 is a carbon-rich nebula, dust is thought to be primarily made up of amorphous carbon.
Based on the best-fit dust model proposed by \cite{Lau2016}, we have applied a similar approach here.
A large grain (LG) component of a single uniform distribution of amorphous carbon \citep{Rouleau1991} with radii {1.5~\micron} is considered.
A log-normal distribution of $(3.1-12)\times10^{-4}$~\micron neutral polycyclic aromatic hydrocarbons (PAHs) centered at $6.4\times10^{-4}$~\micron with a width $\sigma=0.1$ is adopted considering the refractive index data from \cite{Draine2007}.
A power-law grain size distribution, $n(a)\propto a^{-\alpha}$ with index $\alpha = 3.5$, $\rm{a_{min}=12\times10^{-4}}$~\micron, and $\rm{a_{max}=0.4}$~\micron of amorphous carbon \citep{Rouleau1991} very small grains (VSGs) is finally considered.
{Following \cite{Lau2016}, we adjusted the dust properties of three independent components with LGs, VSGs, and PAHs constituting
96.51\%, 1.93\%, and 1.56\% of the total dust by mass, respectively.
With these different components of grain size distribution, we obtain dust-to-gas mass ratio ($M_d/M_g$) of $\sim 1/145$, a ratio of extinction per reddening of $R_V=A_V/E(B-V)=4.118$ and extinction-to-gas ratio = $A_V/N(H) \sim 1.48\times10^{-22}$~mag~cm$^2$.}

\begin{table}
\caption{Physical parameters and corresponding values used in the \textsc{Cloudy} model.}
	\label{tab:phy}
\centering
\resizebox{\linewidth}{!}{\begin{tabular}{lc}
\hline
\hline
{\bf Parameters} & {\bf Values}  \\
\hline
            Geometry & Closed thick shell \\
		Distance & 980 pc ($\sim 3.02\times10^{21}$ cm) \\
		Inner angular radius & 3.1\arcsec ($\sim$ 0.015 pc = $4.55\times10^{16}$ cm) \\
		Str\"{o}mgren angular radius & 4.6\arcsec ($\sim$ 0.022 pc = $6.75\times10^{16}$ cm) \\
            Stellar luminosity & $6.2\times10^3$~L$_{\sun} = 2.37\times10^{37}$~erg~s$^{-1}$ \\
            Stellar effective temperature & $1.98\times10^5$~K \\
            Radiation field & \cite{blac87} + \cite{Draine1996} + CMB \\
            Initial Hydrogen density [$n({\rm H_{tot}})$] & {$2.96\times10^4$ cm$^{-3}$} \\
            Pressure & Constant {($\sim 2.16\times10^9$~K~cm$^{-3}$)} \\
            Mean primary CR ionization rate & $2\times10^{-16}$ s$^{-1}$ \\
            Type of grain & mix of amorphous carbon LG, VSG, and PAH \\
            Dust-to-gas mass ratio & $1/145$ \\
\hline
\end{tabular}}
\end{table}

\begin{table}

\caption{Initial gas-phase elemental abundances with respect to total hydrogen nuclei$^a$ used in the \textsc{Cloudy} model.}
	\label{tab:abn}
\centering
\begin{tabular}{lclc}
\hline
\hline
{\bf Elements} & {\bf Abundance} & {\bf Elements} & {\bf Abundance} \\
\hline
H & 1 & Si & $1.0\times10^{-5}$ \\
He & 0.120 & P & $2.0\times10^{-7}$ \\
C & $7.8\times10^{-4}$ & S & $1.0\times10^{-5}$ \\
N & $1.8\times10^{-4}$ & Cl & $1.7\times10^{-7}$ \\
O & $4.4\times10^{-4}$ & Ar & $2.7\times10^{-6}$ \\
F & $3.0\times10^{-7}$ & K & $1.2\times10^{-7}$ \\
Ne & $1.1\times10^{-4}$ & Ca & $1.2\times10^{-8}$ \\
Na & $1.9\times10^{-6}$ & Fe & $5.0\times10^{-7}$ \\
Mg & $1.6\times10^{-6}$ & Ni & $1.8\times10^{-8}$ \\
Al & $2.7\times10^{-7}$ & & \\
\hline
\end{tabular}
\tablefoot{$^a$ Elemental abundances are taken from \cite{alle83,khro89}.
}
\end{table}

\subsubsection{Chemistry} \label{sec:chemistry}

The equilibrium abundance of HeH$^+$ is dominated by the reactions noted in \autoref{tab:updated_reaction} (see \autoref{sec:app_C}). Our main update with respect to \cite{neuf20} concerns the radiative association (reaction R1) of He$^+$ with H based on new quantum calculations (see \autoref{sec:app_B}).
For the three other reactions (reactions R2 to R4 noted in \autoref{tab:updated_reaction}), our references are the same, but the rate coefficients were fitted using the modified-Arrhenius equation with good accuracy over the temperature range $100-2\times 10^4$~K.
The other chemical reactions related to HeH$^+$ are adopted from \cite{neuf20,das20,Sil2024}.

\begin{figure}
    \includegraphics[width=0.5\textwidth]{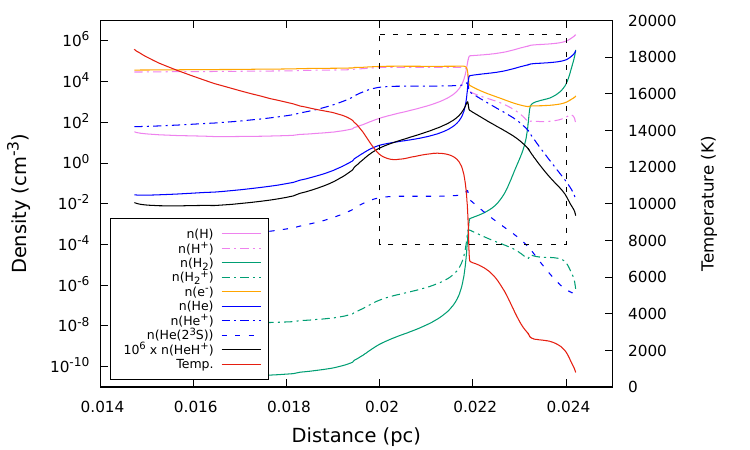}
    \includegraphics[width=0.5\textwidth]{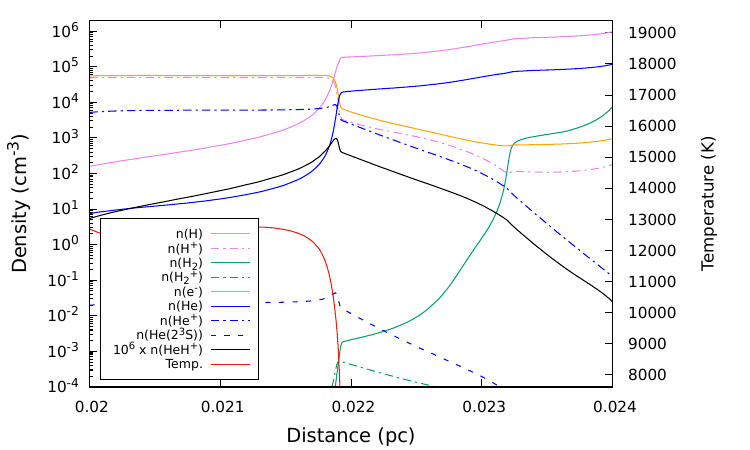}
    \caption{Top panel: Equilibrium temperature and densities of {H, H$^+$, H$_2$, H$_2^+$, e$^-$, He, He$^+$, He($2^3S$), and HeH$^+$} obtained for an isobaric (gas pressure $\sim 2.16\times10^9$~K~cm$^{-3}$) \textsc{Cloudy} model with an initial total hydrogen nuclei density $n(H_{tot}) = 2.61 \times 10^4$~cm$^{-3}$ as function of the distance from the star starting from the illuminated face into the depth of the cloud. Bottom panel: Zoomed-in view of the top panel in the region close to the Str\"{o}mgren radii {(within the black-colored dashed rectangle)}.}
    \label{fig:abundance_density_profile}
\end{figure}

\autoref{fig:abundance_density_profile} shows the equilibrium temperature and density profiles of {H, H$^+$, H$_2$, H$_2^+$, e$^-$, He, He$^+$, He($2^3S$), and HeH$^+$} as a function of the distance from the star starting from the illuminated face (where the starlight striking the cloud) up to a certain depth where the temperature reaches up to 800~K.
The bottom panel of Fig.~\ref{fig:abundance_density_profile} shows a zoomed-in view of the top panel, focusing on the region near the Str\"{o}mgren radius {(within the black-colored dashed rectangle)} where the HeH$^+$ peak appears.
The density peak of HeH$^+$ ($\sim 9.69\times10^{-4}$~cm$^{-3}$) from our model appears close to the helium Str\"{o}mgren radius ($\sim 0.02189$~pc), where half of the He$^+$ has recombined. But it is important to note that the abundance peak of HeH$^+$ ($\sim 1.02\times10^{-8}$) appears closer to the hydrogen Str\"{o}mgren radius ($\sim 0.02186$~pc). The HeH$^+$ peak density is $\sim 2.1$~times lower than that reported in \cite{neuf20}.
It should be noted that the density and temperature profiles shown in Fig.~\ref{fig:abundance_density_profile} slightly differ from \cite{neuf20} (see their Fig.~8) due to the update from \textsc{Cloudy} version c17 to c23, as the microphysics treatment in \textsc{Cloudy} improves over time.
Also, due to the revised physical model parameters (source luminosity and effective temperature, radiation field, grain physics), we had to fix the initial $n({\rm H_{tot}})$ at {$2.96\times10^4$~cm$^{-3}$} for the constant pressure model to obtain the expected position of the Str\"{o}mgren sphere.

\cite{neuf20} have shown that the associative ionization of atomic H with He($2^3S$) is a significant source of HeH$^+$ formation. This finding reflects their update of the rate coefficient to the measurements of \cite{waibel88}, ignored in all previous astrochemical models.
While \textsc{Cloudy} allows us to predict the population of the metastable $2^3S$ level of Helium through its recombination history, and to add it as a reaction partner or product to its chemical network, it is as yet (Version 23.01) unable to consistently couple the underlying chemical reaction network to the statistical equilibrium calculations.
We, therefore, neglect this reaction in our \textsc{Cloudy} modeling. We have, however, tested its impact using a simple steady-state chemical model restricted to the four dominant formation and destruction reactions of HeH$^+$, as identified by \citet{neuf20} (reactions R1, R2, R3, and R4 in \autoref{tab:updated_reaction}), and taking the \textsc{Cloudy} density profiles for H, He$^+$, He(2$^3S$) and e$^-$. We found that the HeH$^+$ peak density is increased by about a factor of 2.3, reaching a peak density of $\sim 2.19\times10^{-3}$~cm$^{-3}$, similar to the results of \cite{neuf20} (see their Fig.~8). The impact of this reaction is thus significant and will be considered in our NLTE \texttt{RADEX} calculations in Sect.~\ref{sec:radex}.

\subsubsection{NLTE calculations} \label{sec:NLTE_cloudy}

\begin{table*}
\caption{Comparison between the observed and \textsc{Cloudy} model predicted line fluxes.}
	\label{tab:line_fluxes_cloudy}
\resizebox{\linewidth}{!}{\begin{tabular}{{lcccccc}}
 \hline
 \hline
{\bf Lines} & {\bf Rest wavelength} & {\bf Observed line flux} & \multicolumn{2}{c}{\bf Predicted line flux ($10^{-18}$~W~m$^{-2}$)} & \multicolumn{2}{c}{\bf Ratio (Observed / Predicted)} \\
  & {\bf (\micron)} & {\bf ($10^{-18}$~W~m$^{-2}$)} & {\bf This work*} & {\bf \cite{neuf20}} & {\bf This work} & {\bf \cite{neuf20}} \\
\hline
		HeH$^+$ $(\upsilon,\ J)=(0, 1 \rightarrow 0,0)$ & 149.091 & $163\pm32$ & {47.35} & 56 & {$3.44\pm 0.68$} & $2.86\pm0.56$ \\
           HeH$^+$ $(\upsilon,\ J)=(1,0 \rightarrow 0,1)\ [P(1)]$ & 3.51532 & $1.55\pm0.16$ & {0.81} & 1.68 & {$1.91 \pm 0.20 $} & $0.92\pm0.095$ \\
            HeH$^+$ $(\upsilon,\ J)=(1,1 \rightarrow 0,2)\ [P(2)]$ & 3.60677 & $2.08\pm0.31$ & {0.91} & 0.89 & {$2.29 \pm 0.34 $} & $2.33\pm0.35$ \\
\hline
        H\,{\sc i} $19-6$ & 3.64493 & $23.9\pm0.22$ & {20.44} & 34.6 & {$1.17 \pm 0.011 $} & $0.69\pm0.006$ \\
        He\,{\sc ii} $13-9$ & 3.54328 & $53.7\pm0.21$ & {44.20} & 108 & {$1.22 \pm 0.005 $} & $0.50\pm0.002$ \\
        He\,{\sc i} $5^3D-4^3P^0$ & 3.70256 & $8.44\pm0.48$ & {11.35} & 12.8 & {$0.74 \pm 0.042 $} & $0.66\pm0.038$ \\
\hline        
\end{tabular}}
\tablefoot{* The line fluxes in \textsc{Cloudy} are given in units of erg~s$^{-1}$~cm$^{-2}$. To compare with the observations, we converted them to W~m$^{-2}$, accounting for the observational angular sizes tabulated in Table~4 (see notes) of \cite{neuf20}.
The conversion formula used are: \\ Predicted flux (W~m$^{-2}$) for HeH$^+$ $(\upsilon,\ J)=(0, 1 \rightarrow 0,0)$ = \textsc{Cloudy} flux (erg~s$^{-1}$~cm$^{-2}$) $\times 10^{-3} \times \frac{1}{4\pi} \times \pi \times (4.6 \times \frac{\pi}{180 \times 3600})^2$. \\
Predicted flux (W~m$^{-2}$) for HeH$^+$ $\upsilon=1 \rightarrow 0 \ P(1), P(2)$, H\,{\sc i}, He\,{\sc ii}, and He\,{\sc i} = \textsc{Cloudy} flux (erg~s$^{-1}$~cm$^{-2}$) $\times 10^{-3} \times \frac{1}{4\pi} \times (0.375 \times 11) \times (\frac{\pi}{180 \times 3600})^2$.}
\end{table*}

In \textsc{Cloudy}, the NLTE radiative transfer is treated in 1D geometry within the escape probability formalism \citep[][and references therein]{ferl17}. We assumed the closed spherically symmetric geometry (implicit for planetary nebulae) with the \texttt{sphere expanding} (default) option so that the Large Velocity Gradient (LVG) or Sobolev approximation was employed.

Spectroscopic data for HeH$^+$, i.e., level energies and Einstein coefficients, were taken from the \texttt{EXOMOL} database \citep{tennyson2016,amaral2019}. This set includes a total of 183 ro-vibrational levels, with 162 bound and 21 quasi-bound states. Because most of the quasi-bound states have large dissociation probabilities, we selected only the 162 bound states, with the highest level $(\upsilon=11, J=1)$ lying 14\,873.7~cm$^{-1}$ (about 21\,400~K) above the ground level. For inelastic collisions with electrons and hydrogen atoms, we have combined several accurate sets of theoretical data using various approximations based on propensity rules, as described in \autoref{sec:app_A}. It should be noted that rate coefficients for electron-impact (de)excitation are typically $3-4$ orders of magnitude larger than those for collisions with hydrogen atoms. In regions where $n(e)/n({\rm H})\gg 10^{-3}$, such as the region close to the Str\"{o}mgren radius in NGC~7027, electron collisions will thus dominate the excitation of HeH$^+$.

In \autoref{tab:line_fluxes_cloudy}, we compare the observed line fluxes reported by \cite{neuf20} for HeH$^+$, H\,{\sc i}, He\,{\sc ii} and He\,{\sc i} to the predictions of our \textsc{Cloudy} model.
The line fluxes in \textsc{Cloudy} are given in units of erg~s$^{-1}$~cm$^{-2}$. To compare with the observations, we converted them to W~m$^{-2}$, accounting for the observational angular sizes {(see notes in \autoref{tab:line_fluxes_cloudy}).
Our model underpredicts the observation by \cite{gust19} for the $J=1\to 0$ pure rotational line flux of HeH$^+$ by a factor of $3.44$, which is similar but slightly worse than the predictions of \cite{neuf20}, primarily due to the missing reaction channel R2 (see Sect.~\ref{sec:chemistry} and \autoref{tab:updated_reaction}) in our \textsc{Cloudy} model.
For both the $P(1)$ and $P(2)$ ro-vibrational transitions, our model underpredicts them by factors of $1.91$ and $2.29$, respectively, while the predictions by \cite{neuf20} for $P(1)$ are better (overpredicted by 1.08), but that for $P(2)$ is similar (underpredicted by 2.33).}
However, the observed $\upsilon=1-0 \ P(2)/P(1)$ line ratio of $1.34\pm0.24$ is consistent with our model prediction (1.12), while the predicted value by \cite{neuf20} is 0.53.
Since this ratio is mainly determined by the relative collisional rate coefficients for ro-vibrational excitation, our closer match with the observations suggests that the collisional propensity rules are best described in our data set. This point is further illustrated in \autoref{sec:app_A}.
Finally, the line-center optical depths of the three HeH$^+$ lines predicted from the \textsc{Cloudy} model are {$6.48\times10^{-2}$, $9.92\times10^{-5}$, and $6.20\times10^{-6}$} for $J=1 \to 0$, $\upsilon=1-0 \ P(1)$, and $P(2)$, respectively.

Our model underestimates the strongest observed recombination lines for H and He$^+$, showing flux discrepancies of about 1.17 and 1.22 times lower than observed, respectively. Conversely, for He, it overestimates the flux by a factor of 1.34. Overall, our model prediction leads to a deviation percentage of $14-34\%$.
The predicted fluxes for one-electron systems (He$^+$ and H), computed from the actual model ionization and temperature structure, were based on the Case B recombination line emissivities presented by \cite{Storey1995}.
For the He recombination line, we utilized the emissivities calculated by \cite{Porter2012,Porter2013}.
Ultimately, our model has a better agreement with the observed data compared to the predictions made by \cite{neuf20}, likely due to the updated physical model parameters and upgrade from \textsc{Cloudy} version c17 to c23.

\begin{table}
\caption{Physical parameters obtained at HeH$^+$ abundance peak from \textsc{Cloudy} model.}
	\label{tab:phy_heh+_peak}
\centering
\resizebox{\linewidth}{!}{\begin{tabular}{lc}
\hline
\hline
{\bf Parameters} & {\bf Values}  \\
\hline
            Kinetic gas temperature, $T_{kin}$ & {10\,380~K} \\
            H atom density, $n({\rm H})$ & {$4.77\times10^4$~cm$^{-3}$} \\
		electron density, $n(e^-)$ & {$4.63\times10^4$~cm$^{-3}$} \\
  HeH$^+$ total column density, $N({\rm HeH^+})$ &  {$9.67\times10^{11}$ [$2.04\times10^{12}$] cm$^{-2}$} \\
\hline
\end{tabular}}
\tablefoot{The HeH$^+$ column density value in the square brackets denotes the estimation from our simplified four-reaction model (see text in Sect.~\ref{sec:radex}).}
\end{table}

\begin{table*}
\caption{Comparison between the observed and \texttt{RADEX} model$^a$ predicted HeH$^+$ line fluxes.}
	\label{tab:line_fluxes_model_C}
\resizebox{\linewidth}{!}{\begin{tabular}{{lccccc}}
 \hline
 \hline
{\bf HeH$^+$ Lines} & {\bf Rest wavelength} & {\bf Upper state} & {\bf Observed line flux} & {\bf Predicted line flux (by \texttt{RADEX})*} & {\bf Ratio} \\
{\bf $(\upsilon^\prime, J^\prime)\to(\upsilon, J)$}  & {\bf (\micron)} & {\bf energy / $k_B$ (K)} & {\bf ($10^{-18}$~W~m$^{-2}$)}  & {\bf ($10^{-18}$~W~m$^{-2}$)} & {\bf (Observed / Predicted)} \\
\hline
		$(0, 1 \rightarrow 0,0)$ & 149.091 & 96.5 & $163\pm32$ & {33.49 [70.43]} & {$4.87 \pm 0.956$ [$2.31\pm 0.454$]} \\
          $(1,0 \rightarrow 0,1)\ [P(1)]$ & 3.51532 & 4188.3 & $1.55\pm0.16$ & {0.57 [1.19]} & {$2.74 \pm 0.282$ [$1.30\pm 0.134$]} \\
            $(1,1 \rightarrow 0,2)\ [P(2)]$ & 3.60677 & 4276.9 & $2.08\pm0.31$ & {0.63 [1.33]} & {$3.31 \pm 0.493$ [$1.57\pm 0.234$]} \\
\hline
\end{tabular}}
\tablefoot{
* \texttt{RADEX} line flux is given erg~s$^{-1}$~cm$^{-2}$. To compare with the observed flux, we convert the \texttt{RADEX} flux unit to W~m$^{-2}$ and accounting for the observational angular sizes (see notes in \autoref{tab:line_fluxes_cloudy}). \\
$^a$ \textsc{Cloudy} parameters: {$T_{kin}=10\,380$~K, $n({\rm e})=4.63\times10^4$~cm$^{-3}$, $n({\rm H})=4.77\times10^4$~cm$^{-3}$, $N[{\rm HeH^+}]=9.67\times10^{11}$~cm$^{-2}$}, line width $=30$~km~s$^{-1}$. \\
The predicted line fluxes and their ratio with observation considering the larger column density of {$2.04 \times 10^{12}$~cm$^{-2}$} are reported in square brackets.}
\end{table*}

\subsection{\texttt{RADEX} modeling and results} \label{sec:radex}

In \texttt{RADEX}, the NLTE radiative transfer is solved for a homogeneous and isothermal medium within the escape probability formalism \citep{vand07}. As with \textsc{Cloudy}, the LVG approximation (option \texttt{expanding sphere}) was used to model the HeH$^+$ emission in NGC~7027.
The \texttt{RADEX} parameters for the physical conditions, namely kinetic temperature $T_{kin}$, atomic hydrogen density $n$(H), and electron density $n$(e$^-$) were extracted from the \textsc{Cloudy} solution at the HeH$^+$ abundance peak, which also coincides with the emission peak for the three detected lines. These parameters are listed in \autoref{tab:phy_heh+_peak} (see also Fig.~\ref{fig:abundance_density_profile}).
The last parameter is the HeH$^+$ column density per unit velocity interval. The HeH$^+$ column density was taken as the \textsc{Cloudy} value (also reported in \autoref{tab:phy_heh+_peak}) and a line width (FWHM) of 30~km~s$^{-1}$ was selected as representative of the three detected lines.
We also considered a larger column density of {$2.04\times 10^{12}$~cm$^{-2}$}, as predicted from our simple four-reaction model where the associative ionization of H with He(2$^3S$) (reaction R2) was included (see \ref{sec:chemistry}).
The background radiation field was assumed to be solely the 2.725~K cosmic microwave background (CMB) radiation field. We checked that the average interstellar radiation field, as implemented in \texttt{RADEX}, has no effect. The nebular and dust continuum emission was also neglected. {We note, however, that radiative pumping due to the internal infrared continuum is not insignificant and should be investigated in future models.}

The predicted line fluxes are reported in \autoref{tab:line_fluxes_model_C}. We can notice that the agreement with the \textsc{Cloudy} results (reported in \autoref{tab:line_fluxes_cloudy}) is very reasonable, with \texttt{RADEX} line fluxes being typically 30\% lower than those from \textsc{Cloudy}. Since the same collisional data set was employed, this modest difference can be attributed to the different NLTE radiative transfer treatments and, in particular, the isothermal and homogeneous assumptions. As with \textsc{Cloudy}, the $\upsilon=1-0 \ P(1)$ and $P(2)$ lines are significantly underpredicted (by a factor of $\sim 3$) while the $\upsilon=1-0 \ P(2)/P(1)$ observational ratio ($1.34\pm0.24$) is well reproduced (1.12). Our predictions using the larger column density of {$2.04\times 10^{12}$~cm$^{-2}$} are also reported in square brackets. {While the line flux computed with \texttt{RADEX} is still too low for the $J=1-0$ pure rotational line}, the agreement with observational values is much better for the two $\upsilon=1-0$ ro-vibrational lines. Overall, the predicted fluxes agree with observations to better than a factor of 2.3. This result supports the importance of the reaction between He($2^3S$) and H as an additional source of HeH$^+$ in NGC~7027.

In order to illustrate the impact of relative collisional rate coefficients on line flux ratios, we investigate in \autoref{sec:app_A} the relation between the $\upsilon=1-0 \ P(2)/P(1)$ line flux ratio ($1.34\pm0.24$) and the ratio of the corresponding fundamental excitation rate coefficients. We show that the two ratios follow a simple linear relationship so that the line flux ratio is very sensitive to the collisional propensity rules. On the other hand, the $\upsilon=1-0 \ P(2)/P(1)$ line flux ratio depends only very weakly on the electron density and temperature (in the HeH$^+$ emission region), as already noted by \cite{neuf20}.

\subsection{\texttt{RADEX/MCMC} modeling and results} \label{sec:mcmc}

In this Section, a second approach where chemical pumping is introduced in the NLTE analysis is used. Indeed, because HeH$^+$ belongs to the class of reactive ions, i.e., species that react fastly with the dominant colliders (here electrons and H atoms), the chemical source and sink terms should be, in principle, included when solving the statistical-equilibrium equations. These two terms describe the so-called chemical (or formation) pumping process, and they were ignored in the previous sections, as well as in \cite{neuf20}. These authors argued that chemical pumping effects should be minor for the detected levels of HeH$^+$ since the destruction rate coefficients are smaller than the excitation rate coefficients. The competition between inelastic and destruction rates is investigated in detail below. In addition, in contrast to the previous NLTE \texttt{RADEX} calculations where the physical conditions and HeH$^+$ column density were taken from the \textsc{Cloudy} results, here \texttt{RADEX} is combined with MCMC sampling in order to determine the \texttt{RADEX} parameters that best reproduce the three lines detected in NGC~7027.

The version of \texttt{RADEX}, as modified by \cite{Faure2017}, was employed in order to include the state-resolved formation and state-resolved destruction rate coefficients of HeH$^+$. The formation of HeH$^+$ was assumed to be entirely dominated by the radiative association of He$^+$ and H, as in our \textsc{Cloudy} chemical model. Accurate state-resolved rate coefficients were computed for this purpose and are described in \autoref{sec:app_B}. Thus, this new data set describes the formation of all 162 bound states of HeH$^+(\upsilon, J)$ at kinetic temperatures from 10 to 20\,000~K. We recall that the associative ionization of neutral H with He(2$^3$S) is another important source of HeH$^+$ in NGC~7027, as shown by \cite{neuf20} and confirmed above. However, to our knowledge, no state-resolved rate coefficients are available for this process, but we note that theoretical state-resolved cross sections at the collision energy $E_{col}=50$~meV can be found in \cite{waibel88}. Extending such calculations to higher collision energy would be necessary to include the associative ionization process as a source term. For the destruction of HeH$^+$ by electrons and hydrogen atoms through dissociative recombination and proton transfer, respectively, we have employed accurate sets of theoretical data with some extrapolations and assumptions, as described in \autoref{sec:app_B}.
It is to be noted that we neglected the HeH$^+$ photodissociation reaction ($\rm{HeH^+ + h\nu \rightarrow He^+ + H}$) in the chemical pumping model since this reaction is negligible in the region of the HeH$^+$ abundance peak \citep[see Fig.~9 of][]{neuf20}.

MCMC is a statistical method used to estimate the parameters of a model by iteratively generating random samples from the prior probability distributions. 
To calculate the likelihood of the parameters given the observed values, we utilized the publicly available MCMC Python implementation \texttt{emcee} developed by \cite{foreman2013}. It should be noted that MCMC will work even if the number of free parameters is larger than the number of data points, which is the case here (four \texttt{RADEX} free parameters versus three observed line fluxes). In this situation, however, some of the parameters will be unconstrained, as described below.

The MCMC analysis is detailed in the \autoref{sec:app_D}. In order to illustrate the convergence of the four \texttt{RADEX} free parameters (kinetic temperature, atomic hydrogen density, electron density, and column density) considering the applied \texttt{RADEX/MCMC} sampling, we present the corner diagram for the achieved solution in Fig.~\ref{fig:corner_plots_radmcmc}. The corner diagrams show the highest probability distributions resulting from the individual histograms of the mentioned free parameters. The best-fit posterior with uncertainties is noted in \autoref{tab:posteerior}.

\begin{table}
\caption{Summary of the best-fit posterior physical parameters obtained from the \texttt{RADEX/MCMC} model with chemical pumping.}
	\label{tab:posteerior}
\centering
\resizebox{\linewidth}{!}{\begin{tabular}{lcc}
\hline
\hline
{\bf Parameters} & \multicolumn{2}{c}{\bf Values}  \\
 & {\bf Logarithmic scale} & {\bf Linear scale} \\
\hline
            & & \\
            Kinetic gas temperature, $T_{kin}$ (K) & {$3.39^{+0.29}_{-0.25}$} & {$2.45^{+2.33}_{-1.07}\times10^3$} \\
            & & \\
            electron density, $n(e^-)$ cm$^{-3}$ & {$5.16^{+0.59}_{-0.83}$}
            & {$1.45^{+4.17}_{-1.24}\times10^5$} \\
            & & \\
            H atom density, $n({\rm H})$ cm$^{-3}$ & {$4.82^{+0.92}_{-1.22}$} & {$6.61^{+48.39}_{-6.21}\times10^4$} \\
            & & \\
		  HeH$^+$ total column density, $N[HeH^+]$ cm$^{-2}$ & {$12.31^{+0.47}_{-0.12}$} & {$2.04^{+3.99}_{-0.49}\times10^{12}$} \\
  & & \\
\hline
\end{tabular}}
\tablefoot{
The uncertainties quoted represent the 16th and 84th percentiles (equivalent to a $1\sigma$ for a Gaussian distribution).
}
\end{table}

\begin{table*}
\caption{Comparison between the observed and the \texttt{RADEX/MCMC} best-fitted model parameters$^a$ predicted HeH$^+$ line fluxes.}
	\label{tab:line_fluxes_model_A}
\resizebox{\linewidth}{!}{\begin{tabular}{{lccccc}}
 \hline
 \hline
{\bf HeH$^+$ Lines} & {\bf Rest wavelength} & {\bf Upper state} & {\bf Observed line flux} & {\bf Predicted line flux (by \texttt{RADEX/MCMC})*} & {\bf Ratio} \\
{\bf $(\upsilon^\prime, J^\prime)\to(\upsilon, J)$}  & {\bf (\micron)} & {\bf energy / $k_B$ (K)} & {\bf ($10^{-18}$~W~m$^{-2}$)}  & {\bf ($10^{-18}$~W~m$^{-2}$)} & {\bf (Observed / Predicted)} \\
\hline
		$(0, 1 \rightarrow 0,0)$ & 149.091 & 96.5 & $163\pm32$ & {171.60} & {$0.95 \pm 0.186 $} \\
          $(1,0 \rightarrow 0,1)\ [P(1)]$ & 3.51532 & 4188.3 & $1.55\pm0.16$ & {1.80} & {$0.86 \pm 0.089 $} \\
            $(1,1 \rightarrow 0,2)\ [P(2)]$ & 3.60677 & 4276.9 & $2.08\pm0.31$ & {1.92} & {$1.08 \pm 0.161 $} \\
\hline
\end{tabular}}
\tablefoot{* \texttt{RADEX} line flux is given erg~s$^{-1}$~cm$^{-2}$. To compare with the observed flux, we convert the \texttt{RADEX} flux unit to W~m$^{-2}$ and accounting for the observational angular sizes (see notes in \autoref{tab:line_fluxes_cloudy}). \\
$^a$ \texttt{RADEX/MCMC} best-fitted model parameters from \autoref{tab:posteerior}: {$T_{kin}=2455$~K, $n(e)=1.45\times10^5$~cm$^{-3}$, $n(H)=6.61\times10^4$~cm$^{-3}$, $N[HeH^+]=2.04\times10^{12}$~cm$^{-2}$}, line width $=30$~km~s$^{-1}$. }
\end{table*}

As shown in \autoref{tab:line_fluxes_model_A}, the agreement between the MCMC best-fit and the observed line fluxes is very good and within observational error bars for the two transitions $(\upsilon,\ J)=(0,1) \to (0,0)$ and $(1,1) \rightarrow (0,2)$. We can notice in \autoref{tab:posteerior} that the best-fit {posterior parameters are overall} in rather good agreement with the values extracted from the \textsc{Cloudy} solution at the HeH$^+$ abundance peak (see Table~\ref{tab:phy_heh+_peak}). However, from the corner plot shown in Fig.~\ref{fig:corner_plots_radmcmc}, we see that {the posterior distributions are not gaussian and clearly bimodal for the electron density. This latter also shows some correlation with the kinetic temperature and the HeH$^+$ column density. Thus, as expected, the small number of detected lines limits the reliability of the parameter estimates. Still, it is interesting to compare the electron density and HeH$^+$ column density inferred from our \texttt{RADEX/MCMC} modeling with the \textsc{Cloudy} results: the HeH$^+$ column density is twice as large while the electron density is larger by a factor of 3. Compared to the four-reaction model, the HeH$^+$ column density is in excellent agreement. These results will be further discussed in Sect.~\ref{sec:discussion} below.}

\begin{figure}[h]
\centering
    \includegraphics[width=0.50\textwidth]{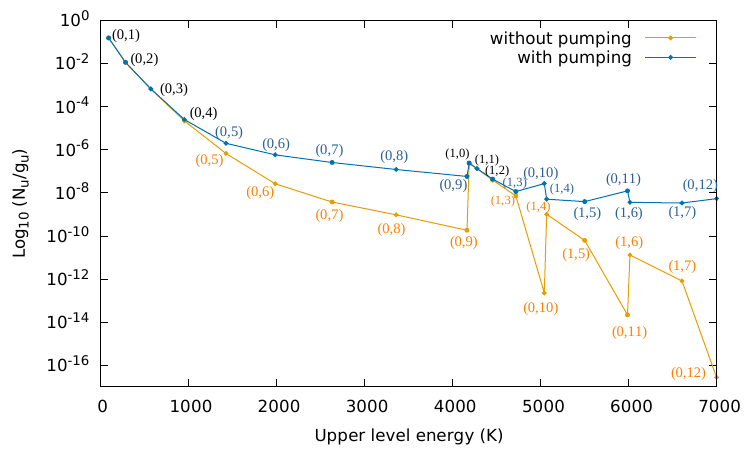}
    \caption{Ro-vibrational population diagram of HeH$^+$~$(\upsilon,\ J)$ as predicted by \texttt{RADEX} for the best-fit parameters noted in \autoref{tab:posteerior}.}
    \label{fig:ro-vib_pop}
\end{figure}

To assess the role of chemical pumping in our \texttt{RADEX} modeling, we have further run an MCMC model where the chemical source and sink terms are ignored in the statistical-equilibrium equations. The best-fit parameters were found to be practically identical to when these terms are included, demonstrating that the three detected lines are not sensitive to the chemical terms. This result confirms the assumption of \cite{neuf20} based on their estimate of destruction and excitation rates, as mentioned above. Indeed, according to our collisional data set, the destruction state-resolved rate coefficients for both electrons and H atoms are (at $10^4$~K) of the order of $10^{-9}$~cm$^{3}$~s$^{-1}$, while the electron-impact excitation rate coefficients from the ground $(\upsilon, J)=(0, 0)$ to the three upper levels $(0, 1)$, $(1, 0)$, and $(1, 1)$ are $4.1\times 10^{-7}$~cm$^{3}$~s$^{-1}$, $3.1\times 10^{-9}$~cm$^{3}$~s$^{-1}$, and $5.3\times 10^{-9}$~cm$^{3}$~s$^{-1}$, respectively.
On the other hand, the electron-impact rate coefficients follow propensity rules, as discussed in \autoref{sec:app_A}, and in practice, all transitions with $\Delta J\geq 4$ or $\Delta \upsilon \geq 2$ have rate coefficients much lower than $10^{-9}$~cm$^{3}$s$^{-1}$.
This is illustrated in Fig.~\ref{fig:ro-vib_pop} where the relative population (divided by the degeneracy) of levels with upper energy lower than 7000~K is plotted as a function of the upper-level energy. The two \texttt{RADEX} models, with and without chemical pumping included, are compared using the parameters listed in \autoref{tab:posteerior}. We can clearly observe that the two models agree for low $J\leq 4$ and that deviations occur for higher and less populated levels, as expected. The population of highly excited levels is thus found to be much larger by up to eight orders of magnitude for the highest plotted level $(\upsilon,\ J)=(0, 12)$ when chemical pumping is taken into consideration. The higher population of excited states reflects their relatively short chemical lifetime and the memory of the HeH$^+$ formation process, as observed previously for CH$^+$ in the same source \citep{Godard2013,Faure2017,neuf21}.

\begin{figure*}[h]
    \includegraphics[width=0.49\textwidth]{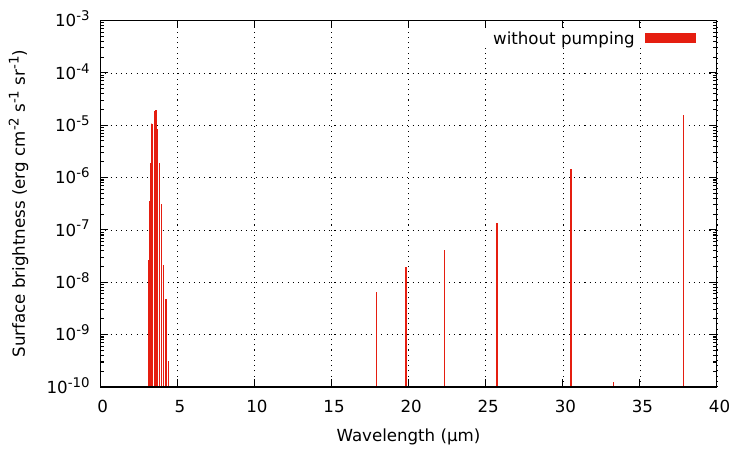}
    \includegraphics[width=0.49\textwidth]{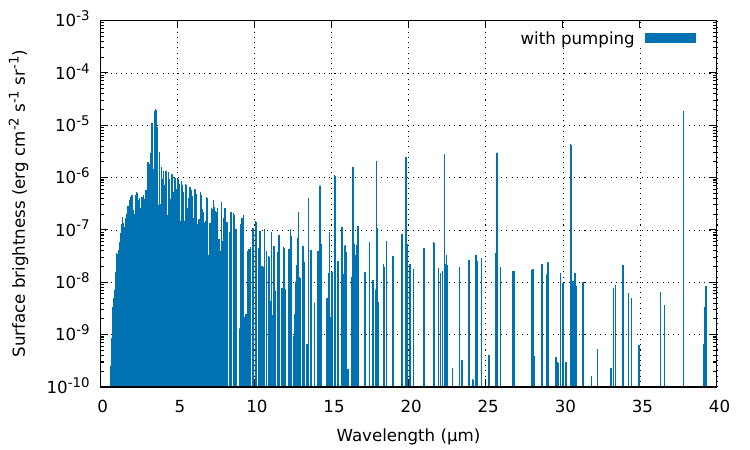}
    \caption{HeH$^+$ line surface brightness variation with wavelength (in \micron) as predicted by the best-fit \texttt{RADEX/MCMC} model parameters noted in \autoref{tab:posteerior}. The left panel is without pumping, and the right panel is considering pumping.}
    \label{fig:model_with_without_pumping}
\end{figure*}

Finally, in order to illustrate how the HeH$^+$ emission spectrum retains some memory of the state-specific formation, we have plotted in Fig.~\ref{fig:model_with_without_pumping} the HeH$^+$ line surface brightness variation with wavelength in the range $0-40$~\micron.
This plot considers the best-fit \texttt{RADEX/MCMC} model parameters mentioned in \autoref{tab:posteerior}.
Furthermore, \autoref{tab:HeH+_transitions} presents the surface brightness of the most intense low-excitation HeH$^+$ transitions with and without pumping based on our best-fit \texttt{RADEX/MCMC} model.
It is evident that the most intense lines (with surface brightness exceeding $10^{-5}$~erg~cm$^{-2}$~s$^{-1}$~sr$^{-1}$), including the three already detected, remain unaffected by pumping. 
Conversely, a significant number of lines exhibit much higher intensities with pumping compared to without, attributed to the substantially elevated populations of their upper levels.
According to our best-fit \texttt{RADEX/MCMC} model with chemical pumping, it is predicted that within the JWST range (0.6~--~28.3~\micron), the next lines to be detected in terms of surface brightness should be the $(\upsilon, J)=(1,1)\to(0,0)$ and $(1,2)\to(0,3)$ lines at 3.36~\micron and 3.71~\micron, respectively.  

\begin{table*} 
\caption{Line surface brightness of the most intense low-excitation pure rotational and ro-vibrational transitions of HeH$^+$ as predicted by the best-fit \texttt{RADEX/MCMC} model parameters noted in \autoref{tab:posteerior}.
\label{tab:HeH+_transitions}}
\centering
\begin{tabular}{ccccc}
\hline
{\bf HeH$^+$ lines} & {\bf $E_U/k_B$} & {\bf Wavelength} & \multicolumn{2}{c}{\bf Surface Brightness (erg~cm$^{-2}$~s$^{-1}$~sr$^{-1}$)} \\
$(\upsilon^\prime, J^\prime)\to(\upsilon, J)$ & {\bf (K)} & {\bf (\micron)} & {\bf Without pumping} & {\bf With pumping} \\
\hline
\multicolumn{5}{c}{\bf Pure rotational lines ($\upsilon=0$)} \\
\hline
$(0,1)\to(0,0)$ & 96.48 & 149.10 & {$1.10\times10^{-4}$} & {$1.10\times10^{-4}$}  \\
$(0,2)\to(0,1)$ & 288.87 & 74.76 & {$2.54\times10^{-4}$} & {$2.56\times10^{-4}$} \\
$(0,3)\to(0,2)$ & 576.08 & 50.08 & {$1.11\times10^{-4}$} & {$1.13\times10^{-4}$}  \\
$(0,4)\to(0,3)$ & 956.44 & 37.82  & {$1.53\times10^{-5}$} & {$1.79\times10^{-5}$} \\
$(0,5)\to(0,4)$ & 1427.79 & 30.52 & {$1.43\times10^{-6}$} & {$4.23\times10^{-6}$}  \\
$(0,6)\to(0,5)$ & 1987.45 & 25.70 & {$1.36\times10^{-7}$} &  {$2.99\times10^{-6}$} \\
$(0,7)\to(0,6)$ & 2632.27 & 22.31 & {$4.06\times10^{-8}$} &  {$2.77\times10^{-6}$} \\
$(0,8)\to(0,7)$ & 3358.67 & 19.80 & {$1.95\times10^{-8}$} &  {$2.46\times10^{-6}$} \\
$(0,9)\to(0,8)$ & 4162.64 & 17.89 & {$6.49\times10^{-9}$} & {$2.03\times10^{-6}$} \\
\hline
\multicolumn{5}{c}{\bf Ro-vibrational lines ($\upsilon = 1 \to 0$)} \\
\hline
$(1,0)\to(0,1)\ [P(1)]$ & 4188.26 & 3.51 & {$1.82\times10^{-5}$} & {$1.86\times10^{-5}$} \\
$(1,1)\to(0,2)\ [P(2)]$ & 4276.91 & 3.60 & {$1.92\times10^{-5}$} & {$1.98\times10^{-5}$} \\
$(1,1)\to(0,0)\ [R(0)]$ & 4276.91 & 3.36 & {$1.07\times10^{-5}$} & {$1.11\times10^{-5}$} \\
$(1,2)\to(0,3)\ [P(3)]$ & 4453.70 & 3.71 & {$8.19\times10^{-6}$} & {$9.12\times10^{-6}$} \\
$(1,2)\to(0,1)\ [R(1)]$ & 4453.70 & 3.30 & {$6.61\times10^{-6}$} & {$7.35\times10^{-6}$} \\
$(1,3)\to(0,4)\ [P(4)]$ & 4717.50 & 3.83 & {$1.90\times10^{-6}$} & {$3.02\times10^{-6}$} \\
$(1,3)\to(0,2)\ [R(2)]$ & 4717.50 & 3.25 & {$1.86\times10^{-6}$} & {$2.95\times10^{-6}$} \\
\hline
\end{tabular}
\end{table*}


\section{Discussion and conclusions} \label{sec:discussion}

In this study, we utilized three different NLTE models (\textsc{Cloudy}, \texttt{RADEX}, and \texttt{RADEX/MCMC} with chemical pumping) to conduct a comprehensive analysis aimed at determining how the HeH$^+$ emissions from NGC~7027 can help to evaluate our understanding of the HeH$^+$ chemistry and excitation. One important finding is that including the reaction of H with He($2^3S$) as an additional source of HeH$^+$ increases its column density by a factor of 2.1 and brings \texttt{RADEX} predictions more in line with observations, with line fluxes that agree within a factor of $\sim$ 2. This result supports the results of \cite{neuf20} and underlines the importance of including He($2^3S$) as a separate species and consistently coupling the resulting chemical population in a future update of \textsc{Cloudy}. This also suggests that our current knowledge of the HeH$^+$ chemistry should be reliable and that it can be employed in other similar environments with good confidence. 
We note in this context that for early Universe chemistry, the formation path to HeH$^+$ is yet another reaction, the radiative association of He with H$^+$, for which accurate rate coefficients are also available \citep{Courtney2021}. 

A perplexing result is that the electron density inferred from our \texttt{RADEX/MCMC} model, {$n(e) = 1.45\times10^5$~cm$^{-3}$, is a factor of $\sim 3$} larger than the electron density predicted by \textsc{Cloudy} at the position of the HeH$^+$ peak, {$n(e) = 4.63\times10^4$~cm$^{-3}$}, which is similar to the average value of $4.7\times10^4$~cm$^{-3}$ yielded by density-sensitive diagnostics from ionic atomic lines \citep{Zhang05}. It is unclear whether HeH$^+$ may serve as a useful probe for electron density in NGC~7027. It may only provide insights into the thin layer where it forms, which is not representative of either the ionized nebula or the surrounding molecular envelope. In any case, a more solid estimate of the electron density will require the identification of additional HeH$^+$ lines, which would also lift the degeneracy with the kinetic temperature and column density.
In this context, we stress that our current results suggest that new ro-vibrational transitions of HeH$^+$ such as $\upsilon = 1-0 \ R(0)$ and $P(3)$ could be detected with the JWST in less than an hour (NIRSpec integral field unit, JWST exposure time calculator version 4.0). Another related important result is that uncertainties in the electron-impact excitation rate coefficients have been largely removed in this study, as highlighted by our analysis of the $\upsilon=1-0 \ P(2)/P(1)$ line ratio.

We have also investigated the impact of the chemical terms on the HeH$^+$ excitation and found that the three detected lines are insensitive to chemical pumping, in contrast to relatively \enquote{short-lived} and less populated levels. As a result, HeH$^+$ chemical pumping effects in NGC~7027 could be probed observationally by detecting lines with surface brightness typically larger than $\sim~10^{-6}$~erg~cm$^{-2}$~s$^{-1}$~sr$^{-1}$.

As a caveat, we note that NGC 7027 is a dynamically young nebula \citep[600~years,][]{mass89,zijl08}, while our modeling assumes steady-state equilibrium. In particular, there may be a significant
mixing of neutral and atomic gas phases. The expansion of the molecular shell can be described by a radial velocity field, $\upsilon_\mathrm{r}=10-13\,\mathrm{km s^{-1}}$ and an additional linear component along its major axis of $6-8\,\mathrm{km s^{-1}}$ and, locally, up to $\sim 40\,\mathrm{km s^{-1}}$ \citep{Santander2012}. The latter high-velocity gas is also seen in ro-vibrational $\mathrm{H_2}$ emissions from the photon-dominated region (PDR) and in the ionized gas \citep[$\mathrm{Br\,\gamma}$ emission,][]{Cox2002}, evidencing a system of jets, or the footprint of a precessing jet \citep{bubl23}. In view of this complexity, some mixing of molecular and ionized gas phases seems inevitable. As a matter of fact, an immersion of molecular knots in the ionized gas of planetary nebulae has been evidenced in later evolutionary stages observationally easier to access, e.g., in the Helix nebula \citep{Meixner2005} or NGC~2828 \citep{Derlopa2024}. \cite{bubl23} indeed locate $\mathrm{CO^+}$ emission in front of the PDR, fully exposed to EUV radiation. In turn, this implies a local coexistence of $\mathrm{H_2^+}$ and He, which could further enhance the $\mathrm{HeH^+}$ abundance through the reversal of the proton-transfer reaction R4 (\autoref{tab:updated_reaction})\footnote{This reaction, $\rm{He + H_2^+ \rightarrow HeH^+ + H}$, is included in our network with a rate coefficient of $\rm{3.00\times10^{-10}~exp(-6717/T)}$~cm$^3$~s$^{-1}$ \citep{blac78}.}. It is possible to provide a rough estimate of whether this path is important: At the maximum $\mathrm{HeH^+}$ density ($T \simeq 10^4$~K), the formation path via reaction $R1$ exceeds that of the reverse of $R4$ by a factor of 750, with a minimum value of 75 at a temperature of 6000~K where the $\mathrm{HeH^+}$ density is an order of magnitude lower. In our {\sc{Cloudy}} model, $\mathrm{H_2^+}$ is produced by photo-ionization of $\mathrm{H_2}$ or charge transfer with $\mathrm{H^+}$ or $\mathrm{He^+}$; only an inflow of molecular gas, as suggested by \cite{Black1983}, could potentially produce the two-order-of-magnitude overabundance of $\mathrm{H_2^+}$ required for the reverse of R4 to compete with R1. Such a scenario is out of the capabilities of our model and reserved for future work.

Finally, we have approximated the elongated nebula as spherically symmetrical to simplify our modeling approach. However, we highlight that considering an ellipsoidal geometry could provide a better prediction of the physical conditions. Investigating how ellipsoidal or more complex geometries affect radiative transfer is a challenging task that requires further study. While addressing this complex geometry is beyond the scope of the present work, it remains an important area for future investigation.


\section*{Data availability}
The new extended HeH$^+$ ro-vibrational inelastic dataset underlying this article will be made available in the Excitation of Molecules and Atoms for Astrophysics (EMAA) database\footnote{\url{https://emaa.osug.fr} and \url{https://dx.doi.org/10.17178/EMAA}}.

\section*{Acknowledgements}
{The authors express their gratitude to the referee, John H. Black, for his invaluable insights and constructive comments, which have substantially enhanced the quality of this work.}
M.S. acknowledges financial support from the European Research Council (consolidated grant COLLEXISM, grant agreement ID: 811363).
The work of R.\v{C}. has been supported by the Czech Science Foundation (Grant No. GACR 21-12598S).
Fabrizio Esposito is acknowledged for sending us the state-resolved rate coefficients for the reaction of HeH$^+$ with H. We are grateful to David A. Neufeld for reading the first version of this paper and making very useful comments. We also thank Thierry Forveille and Joel Kastner for interesting discussions.





\bibliographystyle{aa}
\bibliography{references}



\begin{appendix}

\section{Updated reaction rates} \label{sec:app_C}

The HeH$^+$ chemical network from \textsc{Cloudy} version c23.01 was updated for the three dominant formation and destruction reactions R1, R3, and R4, as listed in \autoref{tab:updated_reaction}. For the radiative association of He$^+$ with H (reaction R1), the thermal rate coefficient computed in the present work (see Sect.~\ref{sec:RA}) was adopted. For the dissociative recombination (DR) of HeH$^+$ (reaction R3), the thermal rate coefficient derived by \cite{novotny2019} was used. For the reaction of HeH$^+$ with H (reaction R4), we employed the thermal quasi-classical trajectory (QCT) rate coefficient computed by \cite{esposito2015}. In addition, the impact of the associative ionization of H with He in the metastable level 2$^3$S (reaction R2) was tested in a simplified four-reaction model, as discussed in Sect.~\ref{sec:chemistry}. For this reaction, a thermal rate coefficient was computed using the experimental cross sections of \cite{waibel88} (see their Fig.~13) extrapolated as $1/E_{col}$ up to 20~eV. For all reactions R1-R4, the rate coefficients were fitted using a standard modified-Arrhenius equation. It should be noted that our fits are reliable over the kinetic temperature range $100-20\,000$~K only, and they should be accurate to within $\sim20\%$. 

\begin{table*}
\centering
\caption{Updated reaction rate coefficient adopted in the chemical model.}\label{tab:updated_reaction}
\begin{tabular}{lccc}
 \hline
 \hline
{\bf Reaction Number} & {\bf Reactions} & {\bf Rate coefficients (cm$^3$~s$^{-1}$)} & {\bf References} \\
\hline
R1 (RA) & $\rm{He^+ + H \rightarrow HeH^+ + h\nu}$ & $\rm{2.32\times10^{-16}~exp(-122/T)}$ &  This work \\
R2 (AI) & $\rm{He(2^3S) + H \rightarrow HeH^+ + e^-}$ & $\rm{2.74\times10^{-10}~(T/300)^{-0.451}~exp(-89.2/T)}$ & \cite{waibel88} \\
R3 (DR) & $\rm{HeH^+ + e^- \rightarrow He + H}$ & $\rm{1.16\times10^{-8}~(T/300)^{-0.908}~exp(-267/T)}$ & \cite{novotny2019} \\
R4 (IN) & $\rm{HeH^+ + H \rightarrow H_2^+ + He}$ & $\rm{1.82\times10^{-9}~exp(-121/T)}$ & \cite{esposito2015} \\
\hline
\end{tabular}
\tablefoot{Fits are reliable over the kinetic temperature range $T=100-20\,000$~K.}
\end{table*}


\section{New HeH$^+$ ro-vibrational data for reactive collisions} \label{sec:app_B}

\subsection{Radiative association of He$^+$ with H} \label{sec:RA}

The electronic potential energy curves for the ground ($X^{1} \Sigma^{+}$) and the excited state ($A^{1} \Sigma^{+}$) of HeH$^+$, along with the dipole moment ($\mu_D$) 
were computed by means of the state-averaged complete active space self-consistent field method followed by a configuration interaction calculation as a function of the internuclear distance using the Molpro software \citep{molpro}, as described in \cite{Loreau2010a}. \\

The partial cross sections of the radiative association reaction were calculated by applying the Fermi golden rule in the zero-density limit as,
\begin{equation}\label{eq:partial_cs}
    \sigma_{\upsilon,J}(E) =  \frac{64 \pi^{5} \nu^{3} p}{3 c^{3} k^{2}} \left[ J M^{2}_{\upsilon,J-1;k,J} + (J + 1) M^{2}_{\upsilon,J+1;k,J} \right] , 
\end{equation}
where $\nu$ represents the frequency of the emitted photon, $k$ denotes the wavenumber of the initial state, $p$ is the probability factor in the initial electronic state,
and $J$ indicates the initial rotational quantum number.  Moreover, The term $M_{\upsilon,J \pm 1;k,J}$ corresponds to the electric transition dipole moment given by, 
\begin{equation}\label{eq:dipole matrix}
    M_{\upsilon,J \pm 1;k,J} = \int_{0}^{\infty}\psi_{\upsilon, J \pm 1}(R) \ \mu_D(R) \ f_{k J}(R) \ dR
\end{equation}
where $f_{k J}(R)$ represents the radial, energy-normalized, continuum wavefunctions which were computed through numerical propagation using the symplectic integrator given by \cite{manolopoulos1995symplectic}; and 
 $\psi_{\upsilon, J \pm 1}(R)$ are the bound ro-vibrational wavefunctions obtained by solving the vibrational Schrodinger equation using the discrete variable representation method described in \cite{colbert1992novel}. Consequently, the total cross section is obtained by summing over all possible vibrational and rotational levels of HeH$^{+}$, $\sigma(E)=\sum_{v,J}\sigma_{vJ}(E)$.\\

The partial cross sections were computed in a collision energy range from $1\times10^{-3}$ to $1\times10^{5}$~cm$^{-1}$ with a denser grid around resonances.

The total or state-to-state rate coefficients as a function of the temperature were computed by assuming the Maxwellian velocity distribution of the initial states, 
and averaging over the normalized energy distribution,
\begin{equation} \label{eq:rate_coefficient}
    \alpha(T) = \left( \frac{8}{\pi \mu} \right)^{\frac{1}{2}} \left( \frac{1}{k_{b} T} \right)^{\frac{3}{2}} \int_{0}^{\infty} E \sigma(E) e^{-E/k_{b} T} dE,
\end{equation}
where $\mu$ is the reduced mass, and $k_{b}$ is the Boltzmann constant. Numerical integration was employed to compute the rate coefficients over a range from 10 to $2\times10^{4}$~K.
The results are in good agreement with previous calculations on the same system \citep{Forrey2020}.

\subsection{Dissociative recombination of HeH$^+$ with electrons} \label{sec:DR}

For the DR of HeH$^+$ with electrons, we have used the MQDT cross sections computed by \cite{curik2020}. HeH$^+$ ro-vibrational levels with $\upsilon=0-2$ and $J=0-2$ were considered. In order to cover kinetic temperatures up to 20\,000~K, cross sections were extrapolated as $1/E_{col}$ from 3.3~eV up to 20~eV. It should be noted that the DR cross sections for HeH$^+(\upsilon=0, J=0-2)$ were checked against an experiment performed at the Cryogenic Storage Ring (CSR) \cite{novotny2019} by computing state-specific DR rate coefficients convolved over the CSR electron beam energy distributions. The good agreement observed between theory and experiment suggests that the state-specific thermal (Maxwell-Boltzmann averaged) rate coefficients computed here should be accurate to within $\sim 30\%$ above 100~K. Within $\upsilon=0-2$, the rate coefficients for $J=2$ were used for all levels with $J\geq 3$. This simple extrapolation should have a modest impact on our modeling since rotational effects become negligible above $\sim 1000$~K (see Fig.~5 in \cite{curik2020}).
Similarly, the rate coefficients for $\upsilon=2, J=2$ were employed for all $J$ levels with $\upsilon\geq 3$. We note in this respect that the DR rate coefficients were found to increase significantly with $\upsilon$, especially below 1000~K where the rate coefficients for $\upsilon=1, 2$ can exceed those for $\upsilon=0$ by more than an order of magnitude. At $10^4$~K, the rate coefficient for the levels $(\upsilon, J)=(0, 0)$, $(1, 0)$ and $(2, 0)$ are $3.9\times 10^{-10}$, $1.3\times 10^{-9}$ and $2.5\times 10^{-9}$~cm$^3$s$^{-1}$, respectively.

\subsection{The reaction of HeH$^+$ with H atoms}

For the reaction of HeH$^+$ with H (which produces H$_2^+$ + He), we have employed the rate coefficients computed by \cite{esposito2015} for (bound) levels $\upsilon=0-5, J=0-23$ and kinetic temperatures up to 15,000~K. This set was computed using the QCT approach, which was found to be in excellent agreement with quantum close-coupling calculations for collisional energies larger than $\sim$ 0.1~eV. The QCT thermal rate coefficient at room temperature (298~K) was also found to be in excellent agreement with the experimental result of \cite{karpas1979} (see Fig.~6 in \cite{esposito2015}). For levels not included in the set of \cite{esposito2015}, i.e. levels with $\upsilon\geq 6$, we have used an average temperature-independent rate coefficient of $1.5\times 10^{-9}$~cm$^3$s$^{-1}$. We note in this respect that rotational and vibrational effects are modest above $\sim 1000$~K with all QCT state-specific rate coefficients lying in the range $1-3\times 10^{-9}$~cm$^3$s$^{-1}$. This range is comparable to the magnitude of the above DR rate coefficients so that hydrogen atoms can compete with electrons in regions where $n(e)/n({\rm H}) \lesssim 1$, which is the observed in our \textsc{cloudy} modeling near the HeH$^+$ abundance peak (see Fig.~\ref{fig:abundance_density_profile}).

\section{New HeH$^+$ ro-vibrational data for inelastic collisions} \label{sec:app_A}

\subsection{Electron collisions}

For inelastic collisions with electrons, we have combined two sets of accurate cross sections: the pure rotational cross sections of \cite{hami16} for levels with $\upsilon=0$ and $J\leq 9$ and the ro-vibrational cross sections of \cite{Curik17} for levels with $\upsilon=0, 1$ and $J\leq 4$. Both sets were computed using the $R$-matrix approach, but the multichannel quantum defect theory (MQDT) used by \cite{Curik17} includes resonances and is more accurate than the adiabatic-nuclei-rotation (ANR) approximation used by \cite{hami16}. In order to cover kinetic temperatures up to 20\,000~K, the two sets of cross sections were extrapolated as $1/E_{col}$ up to 20~eV. In addition, for levels higher than $\upsilon=1, J=4$, the dipolar Coulomb-Born (CB) approximation was employed, as described in \cite{Sil2024} (see their Appendix A). For ions with a significant dipole such as HeH$^+$ \citep[1.66~D;][]{Pavanello2005}, the CB approximation is usually reliable for dipolar transitions, i.e., those with $\Delta J=\pm 1$ and $\Delta \upsilon=0, \pm 1$. For such transitions, the CB cross sections agree very well with those of \cite{hami16}, who applied a CB-closure technique, while they are significantly larger than those of \cite{Curik17}. As discussed by these latter authors, the limited number of partial waves included in their calculations is, in fact, expected to underestimate such cross sections. As a result, for dipolar transitions, we have used the cross sections of \cite{hami16} for levels with $\upsilon=0, J\leq 9$ and the CB cross sections otherwise. The corresponding rotational rate coefficients should be accurate to within $\sim$ 30\%, as observed experimentally for CH$^+$ \citep{kalosi2022} whose dipole is very similar (1.68~D). Future MQDT calculations involving a CB-closure technique, such as those used by \cite{forer2024} for CH$^+$, should confirm this result. Finally, while cross sections for dipole-forbidden transitions with $|\Delta J|\geq 3$ and $|\Delta \upsilon| \geq 2$ can be neglected \citep{Curik17,ayou19}, those with $|\Delta \upsilon|=0, 1$ and $|\Delta J|=0, 2$ have magnitudes similar to those with $|\Delta J|=1$. Thus, for upper levels higher than $\upsilon=1, J=4$, de-excitation rate coefficients for such transitions were set identical to the CB rate coefficients for $|\Delta J|=1$. This simple prescription follows the propensity rules and should be accurate within a factor of 3. We note that the typical magnitude of the largest de-excitation state-to-state rate coefficients at 10$^3$~K is $10^{-6}$~cm$^3$s$^{-1}$ for rotational transitions and  $10^{-8}$~cm$^3$s$^{-1}$ for ro-vibrational transitions.

\subsection{Collision with hydrogen atoms}

For inelastic collisions with hydrogen atoms, we have used the rate coefficients computed by \cite{desr20} for levels with $\upsilon=0$ and $J=0-9$ and kinetic temperatures up to 500~K. This set of data was computed using the quantum close-coupling approach and should be accurate to within $\sim$30\%. A flat temperature dependence of the de-excitation rate coefficients was assumed above 500~K. For levels higher than $\upsilon=0, J=9$, we have followed the propensity rules observed by \cite{desr20} for rotational transitions due to H collisions and by \cite{gianturco2021} for ro-vibrational transitions due to collisions with He atoms: rotational transitions with $|\Delta J|=1, 2$ and ro-vibrational transitions with $|\Delta \upsilon|=1$ and $|\Delta J|\leq 5|$ are favored in these collisions. For levels higher than $(0, 9)$ and transitions obeying the previous propensity rules, temperature-independent de-excitation rate coefficients of $10^{-10}$~cm$^3$s$^{-1}$ and $10^{-11}$~cm$^3$s$^{-1}$ were used, respectively. This crude prescription should be accurate to within a factor of $5-10$. While accurate state-to-state rate coefficients at high temperatures are desirable, we note that in the context of NGC~7027, the excitation of HeH$^+$ is dominated by electron collisions due to their much larger rate coefficients.

\subsection{Impact of propensity rules on the $\upsilon=1-0 \ P(2)/P(1)$ line ratio}

\begin{figure}
    \includegraphics[width=0.5\textwidth]{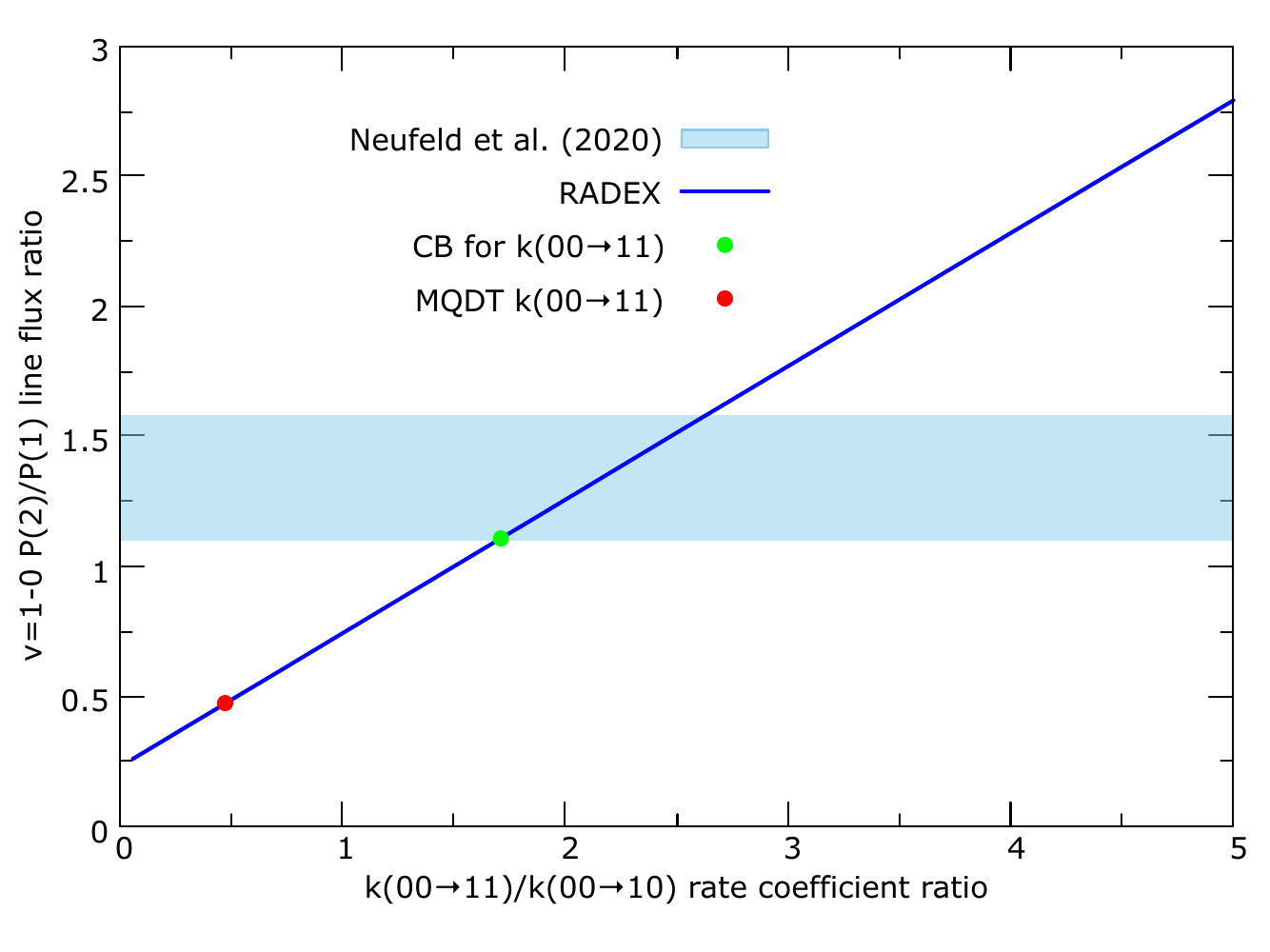}
    \caption{HeH$^+$ line flux ratio predicted by \texttt{RADEX} for the two detected ro-vibrational lines $(\upsilon,\ J)=(1, 1) \rightarrow (0, 2)\ [P(2)]$ and $(1, 0) \rightarrow (0, 1)\ [P(1)]$ as a function of the rate coefficient ratio for the excitation transitions $(0, 0) \to (1, 1)$ and $(0, 0) \to (1, 0)$. The hatched blue area represents the observed ratio. The red and green circles denote the predicted line ratio when using the CB (default) or MQDT rate coefficients for the excitation $(0, 0) \to (1, 1)$. See text for more details.}
    \label{fig:p2p1_line_ratio}
\end{figure}

We have plotted in Fig.~\ref{fig:p2p1_line_ratio} the $\upsilon=1-0 \ P(2)/P(1)$ line flux ratio, as predicted by \texttt{RADEX}, as a function of the rate coefficient ratio for the two fundamental excitations $(0, 0) \to (1, 1) $ and $(0, 0) \to (1, 0)$. At low electron density, line fluxes should be roughly proportional to fundamental excitation rate coefficients. In practice, we have used the same physical conditions as in Sect.~\ref{sec:radex} (those listed in \autoref{tab:line_fluxes_model_C}) with the same collisional data set except that the rate coefficient for the dipolar transition $0, 0 \to 1, 1$ (at 10,500~K) was varied from $2\times 10^{-10}$~cm$^3$s$^{-1}$ to $2\times 10^{-8}$~cm$^3$s$^{-1}$ in order to cover a rate coefficient ratio in the range $0-5$. The default (CB) rate coefficient is $5.4\times 10^{-9}$~cm$^3$s$^{-1}$ while the MQDT value is $1.5\times 10^{-9}$~cm$^3$s$^{-1}$. The rate coefficient for the (dipole-forbidden) transition $0, 0 \to 1, 0$ is kept fixed at the default (MQDT) value of $3.2\times 10^{-9}$~cm$^3$s$^{-1}$. Thus, the rate coefficient ratio in our default collisional data set is 1.7, while it is 0.48 in the original MQDT data set of \cite{Curik17}, as employed by \cite{neuf20}. We can observe in Fig.~\ref{fig:p2p1_line_ratio} that the \texttt{RADEX} predicted line flux ratio is a linear function of the rate coefficient ratio and that this latter should lie in the range $1.7-2.6$ to reproduce the observational line ratio. This result nicely supports the above discussion where CB calculations were preferred to the MQDT calculations of \cite{Curik17} in the case of dipolar transitions.

\section{MCMC fitting details} \label{sec:app_D}

We employed a Bayesian approach to fit three observed HeH$^+$ line fluxes noted in \autoref{tab:line_fluxes_model_A}, to compare models with varying four \texttt{RADEX} parameters listed in \autoref{tab:posteerior}, and then to calculate their probability distributions.
This method of inference involves using prior data to sample posterior distributions.
In practice, we used the MCMC approach to explore parameter space.

A logarithmic likelihood distribution with the formula below was used:
\begin{equation}
    l = \sum_i\left[-\ln(\sigma_i \sqrt{2\pi}) - \frac{1}{2} \frac{(x_i-\mu_i)^2}{\sigma_i^2}\right]
\end{equation}
where x is the observed spectral line flux, $\mu$ is the model line flux, and $\sigma$ is the uncertainty in the observations.

\begin{figure}
    \includegraphics[width=0.5\textwidth]{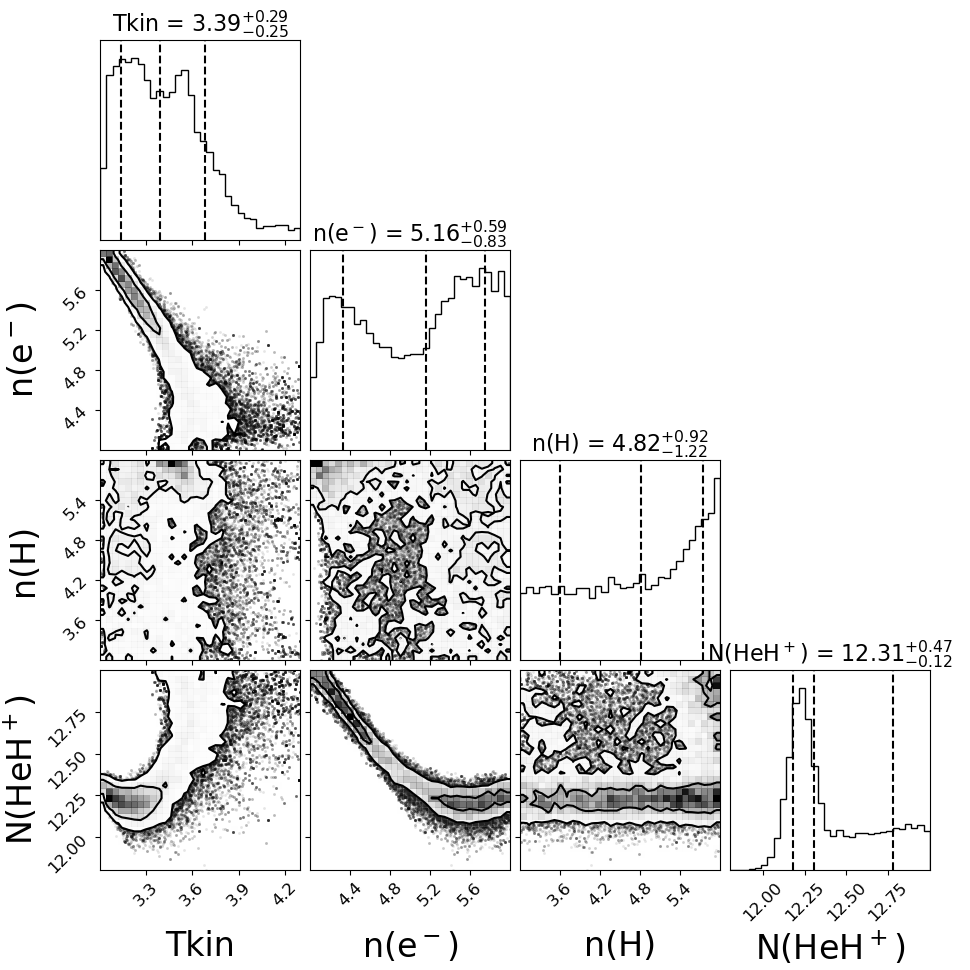}
    \caption{Exploring parameter space using \texttt{RADEX/MCMC} \enquote{corner plot} showing the distribution of $T_{kin}$, $n(e^-)$, $n({\rm H})$, and $N(\rm{HeH^+})$ (in the $\log_{10}$ scale) within the prior intervals [3:4.3], [4:6], [3:6], and [10:13], respectively, in the diagonal panels and the correlation between any pair of them (off-diagonal panels). The central dashed, vertical line in each diagonal panel shows the most likely parameters based on the $\chi^2$ likelihood (the other vertical lines show the 16\% and 84\% quantiles).
    The contours of the corner plot (2D plots) represent the $1\sigma$ and $2\sigma$ levels at the 39.3\% and 86.4\% confidence levels.
}
    \label{fig:corner_plots_radmcmc}
\end{figure}

The corner diagram resulting from the analysis of HeH$^+$ line fluxes is shown in Fig.~\ref{fig:corner_plots_radmcmc}.
The diagonal plots represent the marginal distributions of each parameter, while the off-diagonal plots show the pairwise correlations.
The correlation between $N(\rm{HeH^+})$ and $T_{kin}$ and the electron density suggests that these parameters are closely linked in the model.
For the \texttt{RADEX/MCMC} simulation, we assumed uniform priors spanning the ranges mentioned in the caption of the corner diagram.
It is important to carefully consider the appropriateness of the chosen priors based on prior knowledge or physical constraints.
In our case, we set the initial priors based on the result obtained from the \textsc{Cloudy} model near the peak abundance of HeH$^+$.
The selected initial values were: $T_{kin}=10\,000$~K, $n(e)=5\times10^4$~cm$^{-3}$, $n({\rm H})=5\times10^4$~cm$^{-3}$, and $N(\rm{HeH^+}) = 10^{12}$~cm$^{-2}$.
The MCMC simulation was performed using 48 walkers and 4096 iterations.

\end{appendix}


\end{document}